 \documentclass[preprint]{aastex}
\usepackage{epsfig}
\usepackage{emulateapj5}
\newcommand\etal{et~al.}

\def\LA{Lyman-$\alpha$}

\newcommand{\cotwo}{\mbox{CO$_{2}$}}

\newcommand{\xsig}{$X\,^1\Sigma ^{+}$}
\newcommand{\api}{$A\,^1\Pi$}
\newcommand{\bsig}{$B\,^1\Sigma ^{+}$}
\newcommand{\csig}{$C\,^1\Sigma ^{+}$}
\newcommand{\epi}{$E\,^1\Pi$}

\def\Hone{H\,{\sc i}}
\def\Heone{He\,{\sc i}}
\def\Arone{Ar\,{\sc i}}
\def\Cone{C\,{\sc i}}
\def\Ctwo{C\,{\sc ii}}
\def\Oone{O\,{\sc i}}
\def\Otwo{O\,{\sc ii}}
\def\None{N\,{\sc i}}
\def\lam{$\lambda$}
\def\deg{\hbox{$^\circ$}}

\slugcomment{Accepted for publication in the {\it Astrophysical Journal}}

\shorttitle{FAR-ULTRAVIOLET SPECTROSCOPY OF VENUS AND MARS}
\shortauthors{FELDMAN ET AL.}

\begin{document}

\title{Far-ultraviolet Spectroscopy of Venus and Mars at 4 \AA\
Resolution with the Hopkins Ultraviolet Telescope on Astro-2}

\author{Paul D. Feldman, Eric B. Burgh, Samuel T. Durrance\altaffilmark{1} 
and Arthur F. Davidsen}

\affil{Department of Physics and Astronomy, The Johns Hopkins University\\ 
Charles and 34th Streets, Baltimore, Maryland 21218}

\altaffiltext{1}{Present address: Florida Space Institute, Kennedy Space 
Center, FL 32899}

\begin{abstract}

Far-ultraviolet spectra of Venus and Mars in the range 820~--~1840~\AA\
at $\sim$4 \AA\ resolution were obtained on 13 and 12 March 1995,
respectively, by the Hopkins Ultraviolet Telescope (HUT), which was
part of the Astro-2 observatory on the Space Shuttle {\it Endeavour}.
Longward of 1250 \AA , the spectra of both planets are dominated by
emission of the CO Fourth Positive (\api\ -- \xsig ) band system and
strong \Oone\ and \Cone\ multiplets.  In addition, CO Hopfield-Birge
bands, \bsig\ -- \xsig\ (0,0) at 1151 \AA\ and \csig\ -- \xsig\ (0,0)
at 1088 \AA , are detected for the first time, and there is a weak
indication of the \epi\ -- \xsig\ (0,0) band at 1076 \AA\ in the
spectrum of Venus.  The $B - X$ band is blended with emission from
\Oone\ \lam1152.  Modeling the relative intensities of these bands
suggests that resonance fluorescence of CO is the dominant source of
the emission, as it is for the Fourth Positive system.  Shortward of
Lyman-$\alpha$, other emission features detected include \Otwo\ \lam
834, \Oone\ \lam 989, \Hone\ Lyman-$\beta$, and \None\ \lam\lam 1134
and 1200.  For Venus, the derived disk brightnesses of the \Oone ,
\Otwo , and \Hone\ features are about one-half of those reported by
\citet{hor91} from {\it Galileo} EUV measurements made in February
1990.  This result is consistent with the expected variation from solar
maximum to solar minimum.  The \Arone\ \lam\lam 1048,1066 doublet is
detected only in the spectrum of Mars and the derived mixing ratio of
Ar is of the order of 2\%, consistent with previous determinations.

\end{abstract}


\section{INTRODUCTION}

Ultraviolet observations of the atmospheres of Venus and Mars,
primarily from fly-by or orbiting spacecraft, but also from platforms
above the terrestrial atmosphere, have played an important role in the
study of the composition and structure of the \cotwo\ atmospheres
of these planets.  Early spacecraft experiments were broad-band
photometers that provided information about the spatial distribution of
the emissions in the passband but required remote spectroscopic
measurements to assure the identity of the emitting species.  A
thorough review of this subject is given by \citet{pax92}.  Once the
emitting species were known, narrowband polychromators were developed
and flown on missions such as {\em Mariner 10} \citep{bro74} and {\em
Venera 11} and {\em 12} \citep{ber81}, although the interpretation of
the data from these instruments was far from unambiguous.  With {\em
Mariners 6, 7} and {\em 9}, flown to Mars in the early 1970s, and {\em
Pioneer Venus Orbiter} launched to Venus in 1978, ultraviolet
spectrometers were included in the payloads and provided modest
spectral resolution.  Very little additional spectroscopy, particularly
from newer generations of Earth-orbiting observatories such as the
International Ultraviolet Explorer (IUE) and the Hubble Space Telescope
(HST), has been done and is also summarized by \citeauthor{pax92}.  In
February 1990, {\em Galileo} flew by Venus on its way to Jupiter and
the ultraviolet spectrometers on board made observations that were
reported by \citet{hor91}.

The flight of the Hopkins Ultraviolet Telescope (HUT) on the Astro-2
mission on the Space Shuttle {\it Endeavour} in March 1995 provided an
opportunity to measure the ultraviolet disk spectra of both Venus and
Mars at a spectral resolution ($\sim 4$ \AA) significantly higher than
any of the prior spacecraft observations.  Moreover, since the first
order spectral range of HUT extended to wavelengths as short as 830
\AA, the observations of Venus enabled the resolution of the identity
of the emissions recorded in the narrow-band photometric channels of
the {\it Venera 11} and {\it 12} EUV instruments that had been
interpreted in terms of analogue terrestrial spectra \citep{ber81}.
This paper presents the HUT disk spectra of Venus and Mars together
with the spectral identifications and the disk-averaged brightnesses.
CO fluorescence in both the $B - X$ (0,0) and $C - X$ (0,0)
Birge-Hopfield bands is identified in both Venus and Mars and shown to
be consistent with current models of the CO/\cotwo\ mixing ratios in
the atmospheres of these planets.  Several other emissions are
definitively identified for the first time.

\section{INSTRUMENT and OBSERVATIONS}

The HUT instrument consists of a 0.9-m SiC coated mirror that feeds a
prime focus spectrograph with a photon-counting microchannel plate
detector and photodiode array readout.  Details of the instrument and
its performance and calibration are given by \citet{dav92} and
\citet{kru95,kru99}.  The spectrograph covered 820~--~1840 \AA\ in first
order with a dispersion of 0.51 \AA\ per pixel.  The absolute calibration,
monitored several times during the mission by observations of pure
hydrogen white dwarfs, is considered accurate to better than 5\% at
all wavelengths longward of 912 \AA\ \citep{kru99}.  For the observations
reported here, the $20''$ diameter spectrograph aperture was used, and
was underfilled by both Venus and Mars.  The mean spectral resolution
was determined by the diameter of the planet, 15\farcs1 for Venus and
12\farcs2 for Mars, and the illuminated fraction, giving a resolution
of 4--4.5 \AA\ over the entire spectral band.

At the time of the Astro-2 mission, Venus was at a solar elongation slightly
inside the 45\deg\ solar avoidance constraint of HUT.  Nevertheless, since
the performance of HUT and the supporting Instrument Pointing System was
nominal after ten days of the mission, it was decided to attempt to observe
Venus at a solar elongation of 40\fdg2.  This observation occurred at
UT 05:25 on 13 March 1995 and produced no adverse heating of the telescope
assembly nor appreciably elevated scattered light.  A second observation
was made about 9 hours later.  Both observations were made during orbit day.

\begin{figure}
\begin{center}
\epsfig{file=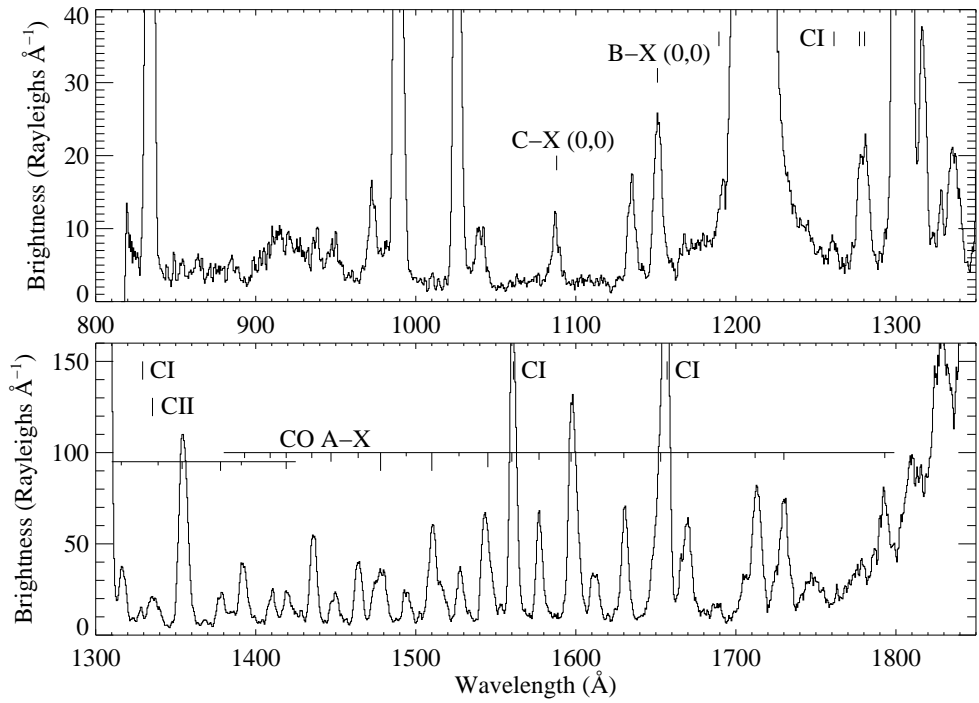, width=5.5in}
\figcaption{HUT spectrum of Venus obtained on 13 March 1995 beginning at
UT 05:31.  The integration time was 1080 seconds and the data have been
smoothed with a running mean over three 0.51 \AA\ wide bins.  Emissions of 
carbon and CO are indicated.  The positions of the strongest solar \LA\ and
\Oone\ \lam1304 pumped CO Fourth Positive bands are indicated separately.
The spectrum also contains emission from the daytime terrestrial
atmosphere. \label{spectv1}}
\end{center}
\end{figure}

\begin{figure}
\begin{center}
\epsfig{file=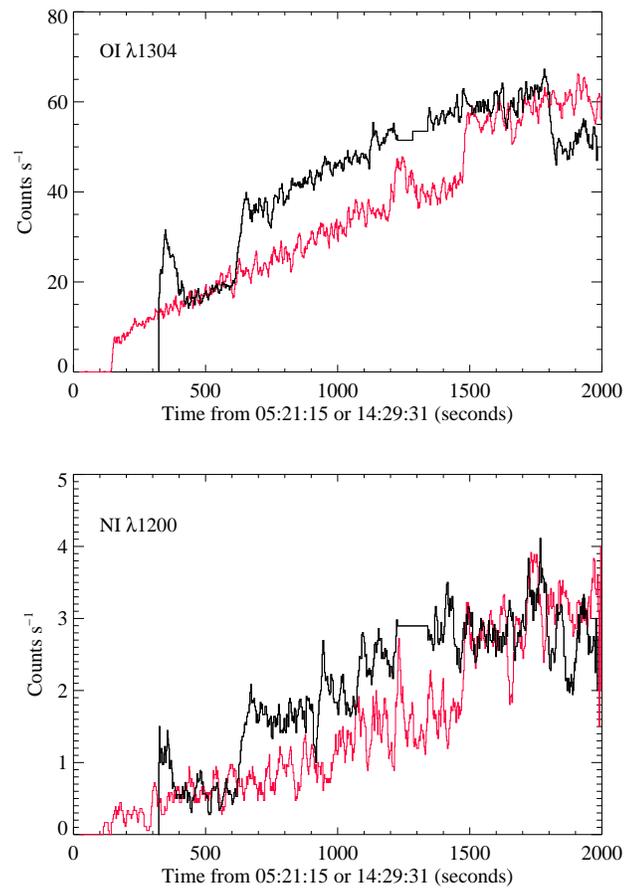, width=3.5in}
\figcaption{Time variation of the \Oone\ \lam1304 and \None\ \lam1200
emissions from the two Venus observations on 13 March 1995.  The black
line begins at UT 05:21 and the red line begins at UT 14:39.  The
separation of terrestrial and Cytherean emissions is clearly
illustrated. \label{time}}
\end{center}
\end{figure}

\begin{figure}
\begin{center}
\epsfig{file=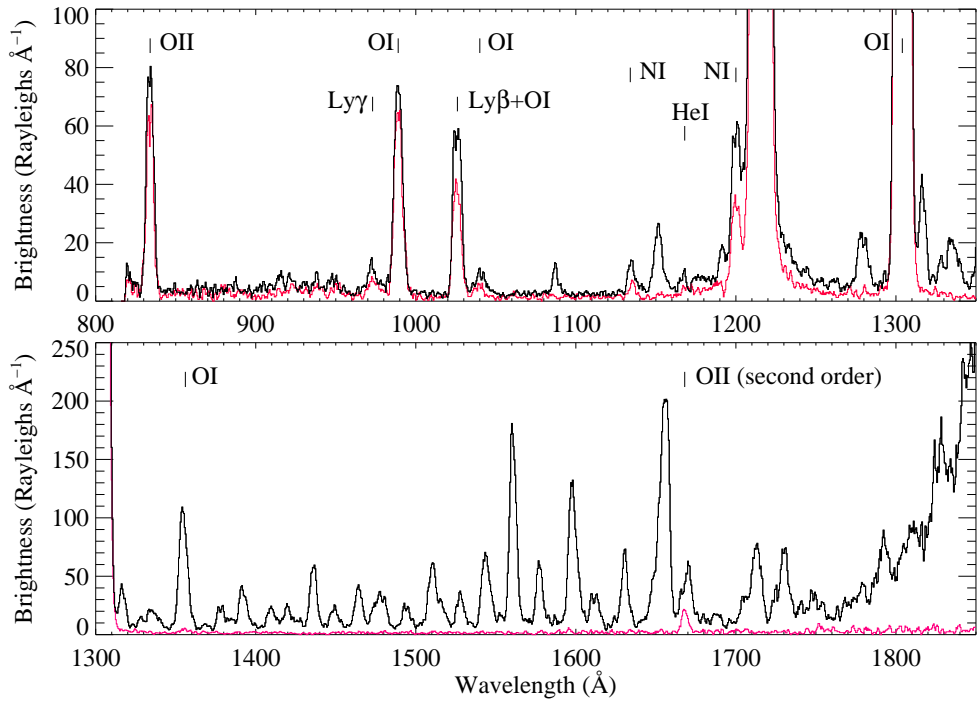, width=5.5in}
\figcaption{Spectra of Venus (with terrestrial background) and
background alone (red line) corresponding to 516 seconds of data
beginning at $\sim$700 seconds in Figure~\ref{time}.  The data have
been smoothed with a running mean over three 0.51 \AA\ wide bins.
Emissions common to both Venus and Earth are indicated. \label{spectv2}}
\end{center}
\end{figure}

A spectrum of Venus from the first observation is shown in
Figure~\ref{spectv1}.  It includes not only the desired disk spectrum
of Venus but also contributions from the terrestrial day airglow.  The
Earth's dayglow includes \Hone\ Lyman-$\alpha$, $\beta$ and $\gamma$,
as well as emissions of atomic and ionic oxygen and atomic nitrogen
that are also present in the dayglow of Venus.  Fortuitously, it took
longer to acquire and lock onto Venus during the second observation so
there are $\sim$500 seconds of background data taken during this
observation that correspond to approximately the same viewing geometry
(with respect to the Earth's atmosphere) as when Venus was in the aperture
during the first observation.  This is illustrated in
Figure~\ref{time}, which shows the count rates for two of the brighter
emissions, \Oone\ \lam1304 and \None\ \lam1200, as a function of time
for both observations.  Figure~\ref{spectv2} shows two 516 second
integrations, Venus + Earth and Earth alone.  The difference spectrum
is then the true spectrum of Venus, and this spectrum is used to derive
the disk brightnesses for the \Oone , \None , and \Hone\ emissions
listed in Table~\ref{tab1}.  For the other emissions, the longer exposure
spectrum of Figure~\ref{spectv1} is used.  The error bars given in
Table~\ref{tab1} are 1-$\sigma$ statistical uncertainties in the observed
counts within a given emission feature.

\placefigure{spectv1}
\placetable{tab1}
\placefigure{time}
\placefigure{spectv2}

\begin{figure}
\begin{center}
\epsfig{file=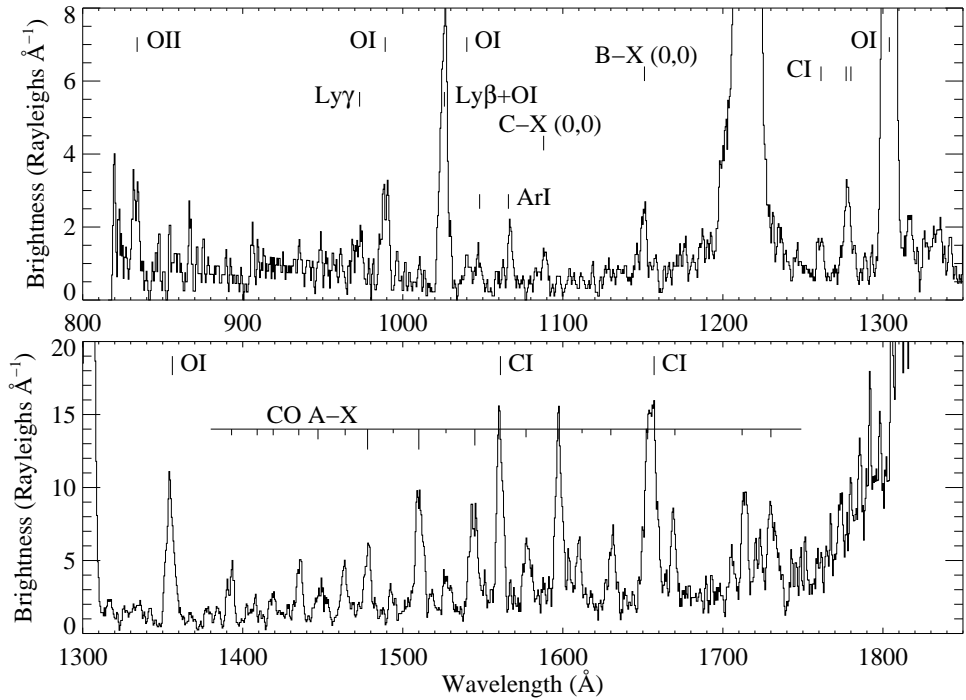, width=5.5in}
\figcaption{HUT spectrum of Mars obtained on 12 March 1995 beginning at
UT 22:33.  The integration time was 1444 seconds and the data have been
smoothed with a running mean over three 0.51 \AA\ wide bins.  This spectrum
was obtained during orbit night and does not show the strong terrestrial
emissions (with the exception of \Hone ) seen in the Venus spectra. 
\label{spectm}}  
\end{center}
\end{figure}

Mars was observed during orbit night beginning at UT 22:33 on 12 March 1995.
The HUT spectrum of Mars is shown in Figure~\ref{spectm}, and the derived
disk brightnesses are also listed in Table~\ref{tab1}.

\placefigure{spectm}

\section{DISCUSSION}

The HUT observations of Venus and Mars represent the highest quality
spectra, in terms of both spectral resolution and instrument sensitivity, obtained
to date for the ultraviolet below 1800 \AA .  The present discussion
will focus on the identification of spectral features and comparison
with previous observations.  Future work will concern the modelling of
these spectra in terms of current atmospheric models based on {\em in
situ} measurements made over the past two decades \citep{pax92}.

\subsection{Carbon Monoxide and Atomic Carbon}

Below 2000 \AA , the ultraviolet dayglow of Venus and Mars is dominated
by emissions of carbon monoxide and carbon \citep{dur81,fox92}.  From
both Fig.~\ref{spectv1} and Fig.~\ref{spectm} we see that a large
number of individual bands of the CO Fourth Positive (\api\ -- \xsig )
system are clearly identified.  When compared with calculated optically
thin fluorescence efficiencies (``g-factors'') for this system
\citep{toz98}, the effect of the twenty-fold increase in the mean
\cotwo\ absorption cross-section from 1700 to 1500 \AA\ \citep{yos96}
is clearly seen
in the enhancement of the weaker longer wavelength bands. 
In addition, the CO is optically thick in that the expected
strong ($v',0$) bands are particularly weak implying that photons are
being pumped out of these bands into other members of the ($v',v''$)
progression by repeated absorption and re-emission.  Bands of the
($14,v''$) and ($9,v''$) progressions, pumped by solar \Hone\ \LA\ and
\Oone\ \lam1304, respectively \citep{kas76,dur81,wol98}, are clearly
seen, particularly the (14,3) band at 1316 \AA\ and the (9,2) band at
1378 \AA, both of which are free of blending by other CO bands or
atomic emissions.  The strong emission feature at $\sim1355$ \AA\ is a
blend of \Oone\ \lam1356 and the CO (14,4) band at 1352 \AA, the
strongest in the solar \LA\ pumped progression, with most of the
observed emission due to CO.

Other observed features include the CO
Hopfield-Birge bands, \bsig\ -- \xsig\ (0,0) at 1151 \AA\ and \csig\ --
\xsig\ (0,0) at 1088 \AA , which are identified for the first time in
the spectra of Venus and Mars.  They
appear to be present in the {\em Galileo} EUV spectrum of Venus \citep{hor91},
but cannot be positively identified at the 30 \AA\ resolution of that
instrument.  The $B - X$ band is blended with emission from
\Oone\ \lam1152.  At first sight, the relative intensities of these
bands and the similarity with low energy electron impact laboratory
spectra \citep{kan95}, suggest that excitation of CO by photoelectrons
is a major source of the emission.  However, as is shown in Section
3.2, it is possible to account for the observed brightness of the $C -
X$ (0,0) band solely on the basis of resonance fluorescence of solar
ultraviolet radiation.  No other bands of either of these systems is
detected, as expected from the strongly diagonal nature of both band
systems and the high pre-dissociation fractions for $v' \geq 1$
\citep{eid91}.  There is also a weak indication of the \epi\ --
\xsig\ (0,0) band at 1076 \AA\ in the spectrum of Venus.
 
In addition to the principal \Cone\ multiplets at 1561 and 1657 \AA, a number 
of other atomic carbon lines are identified in Fig.~\ref{spectv1}.  Some
of these, the multiplets at 1329, 1280, 1277 and 1261 \AA, were
identified in the {\em Mariner 6} and {\em 7} spectra of the upper
atmosphere of Mars \citep{bar71}.  In addition, we identify emission
from \Ctwo\ \lam1335, blended with the (9,1) CO Fourth Positive band
at 1339 \AA, indicating the presence of C$^+$ in the ionospheres of Venus
and Mars.

It is interesting to compare the ratio of these emissions between Venus
and Mars.  From Table~\ref{tab1} we find that for all of the CO bands
and the atomic carbon emissions this ratio is $\approx 15$, whereas the
ratio of solar flux is 5.3.  This had been previously noted by
\citet{moo74}, and reflects the lower CO to \cotwo\ mixing ratio on
Mars compared to that on Venus.

\subsection{The CO Hopfield-Birge Bands}

Emissions from CO shortward of Lyman-$\alpha$ appear in the form of the
Hopfield-Birge bands \csig\ -- \xsig\ (0,0) at 1088 \AA\ and \bsig\ --
\xsig\ (0,0) at 1151 \AA. The $B - X$ band is blended with emission from
\ion{O}{1} $\lambda$1152.  The intensities of these bands are not
consistent with optically thin emission.  Since no other bands of these
systems were detected a scenario of resonant scattering was assumed. The
band is scattered from a column of CO defined by the optical thickness
due to pure absorption by CO$_2$:
\[{\mathcal N}_{CO}=\int^{\infty}_{0}N_{CO}(z)e^{-\tau_{CO_2}(z)}dz\]
and computed using the number densities of the Venus International
Reference Atmosphere (VIRA) model \citep{kea85} and the model of
\citet{fox79} for Mars, with CO$_2$ absorption coefficients taken from
\citet{nak65}.

To test the applicability of these atmospheric profiles to modeling the
CO emission, we compared the column density derived from the (14,3) band
at 1316 \AA\ of the Fourth Positive system (\api\ --
\xsig) and that derived using the above relationship.  The brightness of
the (14,3) band (in rayleighs) in the optically thin limit is \[B = g(14,3)
{\mathcal N}_{CO} \times 10^{-6}\] and measured to be $146\pm25$ R in
the Venus spectrum.  The g-factor for \LA\ absorption in the (14,0)
band was evaluated using the oscillator strength recommended by
\citet{eid99}, $1.80\times10^{-5}$, based on their recent work and the
work of \citet{jol97} and \citet{sta98}, together with the solar
\LA\ line profile from \citet{lem98} normalized to UARS/SOLSTICE
measurements of the solar \LA\ flux \citep{woo96} made at the time of
the {\it Astro-2} mission.  Branching ratios were taken from \citet{kur76}.
We thus derive a CO column density of
$3.4\times10^{16}$ cm$^{-2}$, whereas the VIRA model predicts a column
of $5.2\times10^{16}$ cm$^{-2}$.  The close agreement, considering the
uncertainty in all of the parameters involved in the g-factor
calculation, supports our use of these model atmospheres.

\begin{figure}
\begin{center}
\epsfig{file=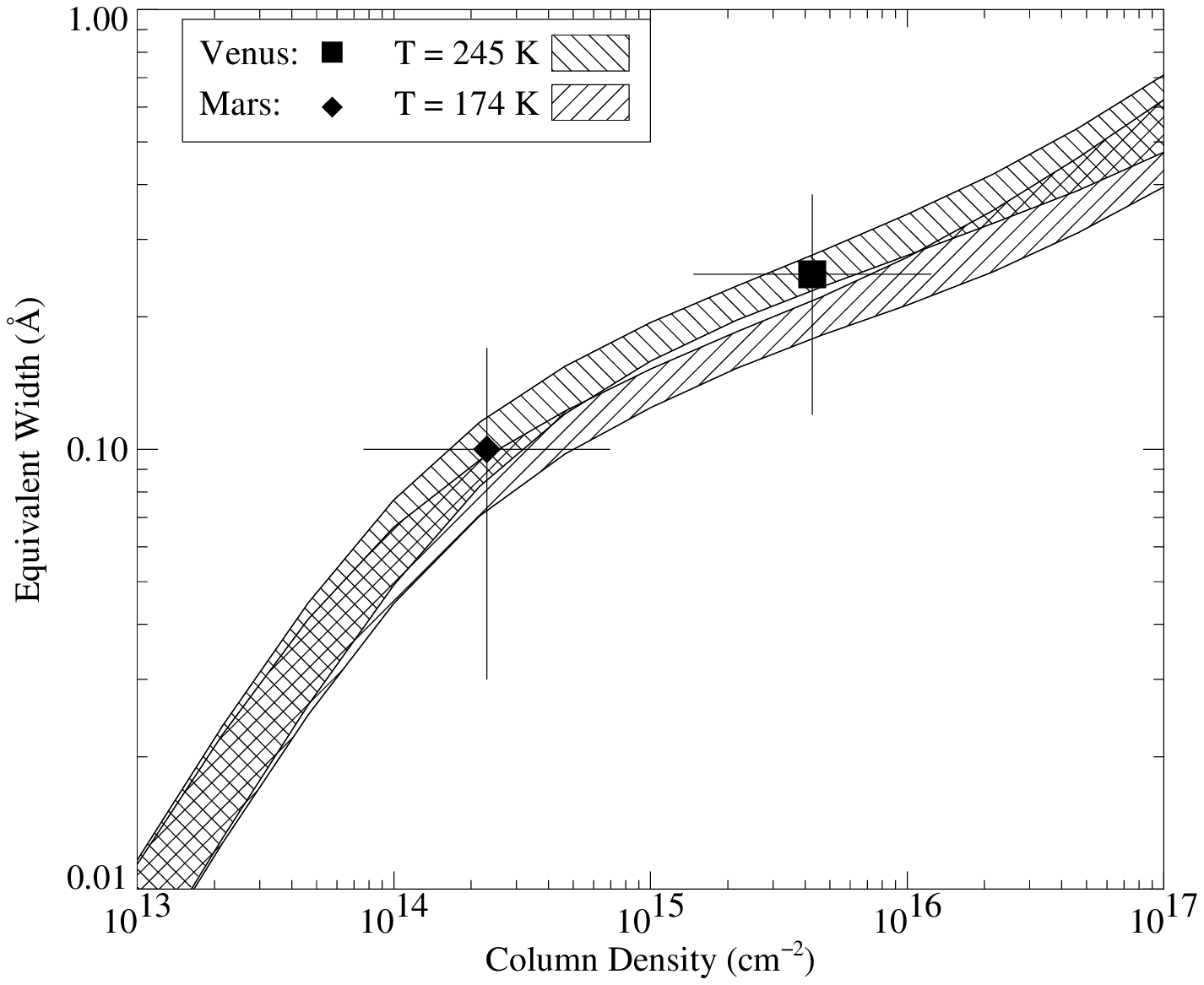, width=3.5in}
\figcaption{Theoretical curves of growth for the \mbox{\csig--\xsig(0,0)} 
transition of CO. The cross-hatched areas represent the bounds of
experimental values for the absorption oscillator strength.  Data
points for Venus and Mars are shown as determined from the HUT data and
atmospheric models from \cite{kea85} and \cite{fox79}.  \label{cog}}
\end{center}
\end{figure}

Following the treatment of \citet{liu96}, a theoretical curve-of-growth
was calculated using the rates of emission from the excited electronic
state of each rovibrational transition to determine the expected
omnidirectional brightness, $4\pi\mathcal{I}$, in rayleighs.  The
equivalent width ($EQW$) of the band is then defined as
$EQW=4\pi\mathcal{I}/\pi\mathcal{F}_\odot$, where $\pi\mathcal{F}_\odot$
is the incident solar flux determined from HUT observations of the moon
and the lunar albedo as given by \citet{hen95}.  To compute the
curve-of-growth, we used a density weighted temperature average to
characterize both the kinetic and rotational temperatures.  Including
transitions through rotational quantum number $J=30$, we produced
curves-of-growth for both Venusian and Martian $C-X$ (0,0) emissions as
shown in Figure~\ref{cog}.  The curves appear as bands with the upper
edge corresponding to an oscillator strength of $1.177\times10^{-1}$
\citep{cha93} as recommended by \citet{mor94}, and the lower to
$6.19\times10^{-2}$ \citep{eid91}.
\placefigure{cog}

The emissions appear to be consistent with resonant scattering from the
upper atmospheres of both planets.  As with the $A-X$ bands, the observed
ratio of the emissions between Venus and Mars is $\approx15$, whereas
the ratio of the solar flux is 5.3.  This difference reflects the lower
CO mixing ratio on Mars as well as a reduced thermospheric temperature
at $\tau_{CO_2}=1$.  Curves of growth for $B-X$ (0,0)
($f=6.5\times10^{-3}$, \citealt{sta99}) were determined in the same manner
as for $C-X$ (0,0) and the equivalent width corresponding to the
appropriate column density was used to calculate an expected $28\pm10$
and $1.8\pm0.8$ R of emission from Venus and Mars respectively.
Subtracting these from the total observed emission results in
\ion{O}{1} $\lambda$1152 emission of about $100\pm14$ R for Venus and
$4.5\pm2.0$ R for Mars, most likely due to direct electron impact
excitation of atomic oxygen.


\subsection{Argon and Helium}

Argon emission at 1048 and 1066 \AA\ is detected only in the spectrum
of Mars (Fig.~\ref{spectm}).  The upper limits for Venus are consistent
with the known argon mixing ratio of $<100$ ppm.  Note that
\Arone\ \lam1066 appears brighter than \Arone\ \lam1048, even though
the fluorescence efficiency of the 1048 \AA\ line is 3.5 times higher
than that of the 1066 \AA\ line.\footnote{The \Arone\ g-factors are
evaluated using solar fluxes inferred from the HUT observations of
sunlight reflected from the Moon \citep{hen95}} This is the result of
the higher \cotwo\ absorption cross-section at 1048 \AA .  Using the
approach of \citet{ber81}, we find an argon mixing ratio of 0.016 and
0.021, from the brightness of the 1048 and 1066 \AA\ lines,
respectively.  Considering the uncertainties in both the data and the
fluorescence efficiencies, this result is in good agreement with the
known value of 0.016 for Mars \citep{bar85}.  The \Arone\ multiplets
near 867 \AA\ are not detected on either Mars or Venus.

Figure~\ref{spectv2} also shows the presence of \Heone\ \lam584 in
second order at 1168 \AA\ in the spectrum of Venus.  The second order
effective area of HUT was only $\sim 1$ cm$^2$, compared to the peak
first order effective area of 25 cm$^2$ near 1150 \AA\ \citep{kru99},
so that the derived brightness has a large statistical uncertainty,
$300\pm200$ R.  Nevertheless, this result is consistent with prior
measurements of \Heone\ emission from {\em Venera} and {\em Galileo}.

\subsection{Other Emissions}

Shortward of Lyman-$\alpha$, other emission features detected include
\Otwo\ \lam 834, \Oone\ \lam 989, \Hone\ Lyman-$\beta$, and
\None\ \lam\lam 1134 and 1200.  The atomic nitrogen features are
identified for the first time, although, as above, \None\ \lam1134
appears to be present in the {\em Galileo} EUV spectrum of Venus.  The disk
brightnesses for both Venus and Mars are listed in Table~\ref{tab1}.
While there are many weak features present in the spectrum of Venus
between 840 and 1000 \AA, most of these are of terrestrial origin,
primarily of atomic and singly ionized nitrogen, and the contributions
from Venus are at least an order of magnitude lower than those reported
by \citet{ste96} from a sounding rocket observation.

\subsection{Comparison with Venera 11/12 and Galileo}

Table~\ref{tab2} gives a comparison of the HUT measurements of the disk
of Venus with those reported from {\em Venera 11} and {\em 12}
\citep{ber81} and, more recently, {\em Galileo} \citep{hor91}.  The
{\em Galileo} flyby of Venus was in February 1990, during solar
maximum, while the HUT observations in March 1995 were at a time
approaching solar minimum.  This is reflected in the roughly factor of
two difference between the two sets of data, which must be considered
excellent agreement.  The {\em Venera 11/12} measurements were made in
December 1978, also approaching solar maximum.  Thus, a similar
argument accounts for the difference in \Oone\ and \Otwo\ brightnesses
between HUT and {\em Venera}.  However, it appears that the {\em
Venera} channels at wavelengths longward of 1500 \AA\ were severely
contaminated by scattered light.
\placetable{tab2}

\section{CONCLUSION}

We have obtained far-ultraviolet spectra of Venus and Mars in the range
820~--~1840~\AA\ at $\sim$4 \AA\ resolution with the Hopkins
Ultraviolet Telescope (HUT) during the {\em Astro-2} mission in March 1995.
The spectra of both planets are rich in CO band emission, some of the
systems being identified for the first time, together with strong
\Oone\ and \Cone\ multiplets.  Resonance fluorescence is identified as
the dominant source of the CO emission.  Atomic nitrogen emissions are
also identified in the spectrum of Venus, while the \Arone\ doublet is
seen only in the spectrum of Mars.  These spectra, obtained at higher
spectral resolution than was possible from earlier fly-by and orbiting
missions to these planets, elucidates and extends the earlier
spectroscopic measurements and should provide guidance in the
interpretation of the far-ultraviolet spectra obtained during the
recent {\em Cassini} fly-by of Venus \citep{ste99}.

\acknowledgments

It is a pleasure to thank the Spacelab Operations Support Group at
Marshall Space Flight Center for their support during the {\em Astro-2}
mission.  We also thank our many colleagues at the Johns Hopkins
University and the Applied Physics Laboratory for their contributions
to the success of the Hopkins Ultraviolet Telescope.  This work was
supported by NASA grant NAG5-5122 and contract NAS5-27000 to the Johns Hopkins
University.




\begin{table}
\begin{center}
\caption{Disk Brightnesses of Venus and Mars.  \label{tab1}}
\medskip
\begin{tabular}{@{}lccc@{}}
\tableline\tableline
Species  & Wavelength & Venus & Mars \\
 & (\AA )     & (rayleighs)  & (rayleighs)  \\
\tableline
\Otwo\  		& 834	& $91\pm41$  	& $7\pm4$ \\
\Arone\ 		& 867	& $<4.0$\tablenotemark{b} 	& $<1.7\tablenotemark{b}$ \\
\Hone\ Lyman-$\gamma$ 	& 973	& $25\pm18$ 	& $3.6\pm2.0$\tablenotemark{a} \\
\Oone\  		& 989 	& $45\pm33$	& $9.4\pm2.6$ \\
\Hone\ Lyman-$\beta$
 + \Oone\  		& 1026 	& $115\pm23$	& $36\pm5$\tablenotemark{a} \\
\Oone\  		& 1040 	& $21\pm7$	& $3.7\pm1.1$ \\
\Arone\  		& 1048 	& $<1.9$\tablenotemark{b} 	& $2.4\pm1.0$ \\
\Arone\  		& 1066 	& $<2.4$\tablenotemark{b} 	& $5\pm2$ \\
\None\  		& 1134	& $35\pm11$	& $\sim1.4$ \\
\None\  		& 1200 	& $77\pm16$	& $\sim12$ \\
\Oone\  		& 1304 	& $2800\pm110$	& $280\pm10$ \\
\Oone\  		& 1356 	& $605\pm28$\tablenotemark{c} 	& $50\pm5$\tablenotemark{c} \\
\Cone\  		& 1561 	& $800\pm27$\tablenotemark{c}	& $57\pm7$\tablenotemark{c} \\
\Cone\  		& 1657 	& $1500\pm50$\tablenotemark{c}	& $110\pm30$\tablenotemark{c} \\

\tableline
CO $C - X$ (0,0)  	& 1088 	& $44\pm6$	& $3.4\pm1.7$ \\
CO $B - X$ (0,0)
 + \Oone\  		& 1152 	& $128\pm10$ 	& $6.3\pm1.9$ \\
CO $A - X$ (0,1)  	& 1597 	& $754\pm30$ 	& $52\pm8$ \\
\tableline
\end{tabular}
\end{center}
\tablenotetext{a}{Includes terrestrial nightglow contribution.} 
\tablenotetext{b}{1-$\sigma$ statistical limit.} 
\tablenotetext{c}{Includes blended CO $A - X$ band.} 
\end{table}


\begin{table}
\begin{center}
\caption{Comparison of HUT Observations of Venus with {\it Venera 11/12}
and {\it Galileo}. \label{tab2}}
\medskip
\begin{tabular}{@{}lcccc@{}}
\tableline\tableline
Species  & Wavelength & HUT & {\it Venera 11/12}\tablenotemark{a} & {\it Galileo}\tablenotemark{b} \\
 & (\AA )     & (rayleighs)  & (rayleighs) & (rayleighs) \\
\tableline
\Heone\  & 584 & $300 \pm 200$  & 275 & $200 \pm 60$ \\
\Otwo\  & 834 & $91\pm41$  & 156 & $180 \pm 60$ \\
\Arone\ & 867 & $<4.0$\tablenotemark{c} & $<55$ & \\
\Oone\  & 989 & $45\pm33$ & & $130 \pm 30$ \\
\Hone\ Lyman-$\beta$ + \Oone\  & 1026 & $115\pm23$ & & $270 \pm 60$ \\
\Arone\  & 1048 & $<1.9$\tablenotemark{c} & $<130$ & \\
\Oone\  & 1304 & $2800\pm110$ & 6200 & \\
CO $A - X$  & 1500\tablenotemark{d} & $965\pm110$ & 2400 & \\
\Cone\  & 1657 & $1500\pm50$ & 12500 & \\
\tableline
\end{tabular}
\end{center}
\tablenotetext{a}{\citet{ber81}.  Average of both missions.} 
\tablenotetext{b}{\citet{hor91}.} 
\tablenotetext{c}{1-$\sigma$ statistical limit.} 
\tablenotetext{d}{45 \AA\ wide band centered at 1500 \AA\ includes
(1,0) and (2,0) bands.} 
\end{table}



\clearpage


\begin{thebibliography}{}

\bibitem[Barth(1985)]{bar85} Barth, C. A. 1985, in {\em The
Photochemistry of Atmospheres}, ed. J. S.  Levine, Orlando: Academic
Press, 337

\bibitem[Barth et al.(1971)]{bar71} Barth, C. A., Hord, C. W.,
Pearce, J. B., Kelly, K. K., Anderson, G. P., and Stewart, A. I. 1971,
\jgr, 76, 2213

\bibitem[Bertaux et al.(1981)]{ber81} Bertaux, J. L., Blamont, J. E.,
Lepine, V. M., Kurt, V. G., Romanova, N. N., \& Smirnov, A. S. 1981,
Planet. Space Sci., 29, 149

\bibitem[Broadfoot et al.(1974)]{bro74} Broadfoot, A. L., Kumar, S.,
Belton, M. J. S., and McElroy, M. B. 1974, Science, 183, 1315

\bibitem[Chan et al.(1993)]{cha93} Chan, W. F., Cooper, G., \&
Brion, C. E. 1993, \jcp, 170, 123

\bibitem[Davidsen et al.(1992)]{dav92} Davidsen, A. F. \etal\ 1992,
\apj, 392, 264

\bibitem[Durrance(1981)]{dur81} Durrance, S. T. 1981, \jgr, 86, 9115

\bibitem[Eidelsberg et al.(1991)]{eid91} Eidelsberg, M., Benayoun, J. J.,
Viala, Y., \& Rostas, F. 1991, \aaps, 90, 231

\bibitem[Eidelsberg et al.(1999)]{eid99} Eidelsberg, M., Jolly, A.
Lemaire, J. L., Tchang-Brillet, W.-UL., Breton, J., \& Rostas, F. 1991, 
\aap, 346, 705

\bibitem[Fox(1992)]{fox92} Fox, J. L. 1992, in {\em Venus and Mars:
Atmospheres, Ionospheres, and Solar Wind Interactions}, ed. J. G.
Luhmann, M. Tatrallyay and R. O.  Pepin, Geophysical Monograph 66,
Washington: AGU, 191

\bibitem[Fox \& Dalgarno(1979)]{fox79} Fox, J. L., \& Dalgarno, A.
1979, \jgr, 84, 7315

\bibitem[Henry et al.(1995)]{hen95} Henry, R. C., Feldman, P. D., Kruk,
J. W., Davidsen, A. F., \& Durrance, S. T. 1995, \apjl, 454, L69

\bibitem[Hord et al.(1991)]{hor91} Hord, C. W., \etal\ 1991, Science, 253, 1548

\bibitem[Jolly et al.(1997)]{jol97} Jolly, A., Lemaire, J. L.,
Belle-Oudry, D., Edwards, S., Malmasson, D., Vient, A., \& Rostas, F.
1997, J. Phys. B, 30, 4315

\bibitem[Kanik et al.(1995)]{kan95} Kanik, I., James, G. K., \& Ajello, J. M. 1995, \pra, 51, 2067

\bibitem[Kassal(1976)]{kas76} Kassal, T. T. 1976, \jgr, 81, 1411

\bibitem[Keating et al.(1985)]{kea85} Keating, G. M., Bertaux, J. L.,
Bougher, S. W., \& Dickinson, R. E. 1985 {\em Adv. Space Res.}, 5, 117

\bibitem[Kruk et al.(1995)]{kru95} Kruk, J. W., Durrance, S. T., Kriss,
G. A., Davidsen, A. F., Blair, W. P., Espey, B. R., \& Finley, D. S.
1995, \apj, 454, L1

\bibitem[Kruk et al.(1999)]{kru99} Kruk, J. W., Brown, T. M., Davidsen, A. F., 
Espey, B. R., Finley, D. S., \& Kriss, G. A., 1999, \apjs, 122, 299

\bibitem[Kurucz(1976)]{kur76} Kurucz, R. L.
1976, Smithsonian Astrophysical Observatory Special Report 374

\bibitem[Lemaire et al.(1998)]{lem98} Lemaire, P., Emerich, C., Curdt, W.,
Sch\"{u}hle, U., and Wilhelm, K. 1998, 
\aap, 334, 1095

\bibitem[Liu \& Dalgarno(1996)]{liu96} Liu, W. \& Dalgarno, A. 1996, \apj, 462, 502

\bibitem[Moos(1974)]{moo74} Moos, H. W. 1974, \jgr, 79, 685

\bibitem[Morton \& Noreau(1994)]{mor94} Morton, D. C. \& Noreau, L.
1994, \apjs, 95, 301

\bibitem[Nakata et al.(1965)]{nak65} Nakata, R. S., Watanabe, K., \&
Matsunaga, F. M. 1965, {\em Sci. Light}, 14, 54

\bibitem[Paxton \& Anderson(1992)]{pax92} Paxton, L. J. \& Anderson, D.
E. 1992, in {\em Venus and Mars: Atmospheres, Ionospheres, and Solar
Wind Interactions}, ed. J. G. Luhmann, M. Tatrallyay and R. O.  Pepin,
Geophysical Monograph 66, Washington: AGU, 113

\bibitem[Stark et al.(1998)]{sta98} Stark, G., Lewis, B. R., Gibson, S. T.,
\& England, J. P., 1998, \apj, 505, 452

\bibitem[Stark et al.(1999)]{sta99} Stark, G., Lewis, B. R., Gibson, S. T.,
\& England, J. P., 1999, \apj, 520, 732

\bibitem[Stern et al.(1996)]{ste96} Stern, S. A. \etal\ 1996,
Icarus, 122, 200

\bibitem[Stewart et al.(1999)]{ste99} Stewart, A. I. F. \etal\ 1999,
\baas, 31, 1174

\bibitem[Tozzi et al.(1998)]{toz98} Tozzi, G. P., Feldman, P. D.,
and Festou, M. C. 1998, \aap, 330, 753

\bibitem[Wolven \& Feldman(1998)]{wol98} Wolven, B. C. \& Feldman, P. D.
1998, in {\em The Scientific Impact of the Goddard High Resolution 
Spectrograph}, ed. J. C. Brandt, T. B. Ake, III \& C. C. Petersen, ASP 
Conference Series 143, San Francisco: ASP, 373

\bibitem[Woods et al.(1996)]{woo96} Woods, T. N., et al.~1996, \jgr, 101, 9541

\bibitem[Yoshino et al.(1996)]{yos96} Yoshino, K, Esmond, J. R., Sun, Y.,
Parkinson, W. T., Ito, K., and Matsui, T. 1996, JQSRT, 55, 53

\end{thebibliography}
\end{document}